\documentstyle[astrop-bib,times]{mn}
\def\preprint{1}		
\def\PsfigVersion{1.9}
\ifx\undefined\psfig\else\endinput\fi

\let\LaTeXAtSign=\@
\let\@=\relax
\edef\psfigRestoreAt{\catcode`\@=\number\catcode`@\relax}
\catcode`\@=11\relax
\newwrite\@unused
\def\ps@typeout#1{{\let\protect\string\immediate\write\@unused{#1}}}
\ps@typeout{psfig/tex \PsfigVersion}

\def\figurepath{./}

\def\@nnil{\@nil}
\def\@empty{}
\def\@psdonoop#1\@@#2#3{}
\def\@psdo#1:=#2\do#3{\edef\@psdotmp{#2}\ifx\@psdotmp\@empty \else
    \expandafter\@psdoloop#2,\@nil,\@nil\@@#1{#3}\fi}
\def\@psdoloop#1,#2,#3\@@#4#5{\def#4{#1}\ifx #4\@nnil \else
       #5\def#4{#2}\ifx #4\@nnil \else#5\@ipsdoloop #3\@@#4{#5}\fi\fi}
\def\@ipsdoloop#1,#2\@@#3#4{\def#3{#1}\ifx #3\@nnil 
       \let\@nextwhile=\@psdonoop \else
      #4\relax\let\@nextwhile=\@ipsdoloop\fi\@nextwhile#2\@@#3{#4}}
\def\@tpsdo#1:=#2\do#3{\xdef\@psdotmp{#2}\ifx\@psdotmp\@empty \else
    \@tpsdoloop#2\@nil\@nil\@@#1{#3}\fi}
\def\@tpsdoloop#1#2\@@#3#4{\def#3{#1}\ifx #3\@nnil 
       \let\@nextwhile=\@psdonoop \else
      #4\relax\let\@nextwhile=\@tpsdoloop\fi\@nextwhile#2\@@#3{#4}}
\ifx\undefined\fbox
\newdimen\fboxrule
\newdimen\fboxsep
\newdimen\ps@tempdima
\newbox\ps@tempboxa
\long\def\fbox#1{\leavevmode\setbox\ps@tempboxa\hbox{#1}\ps@tempdima\fboxrule
    \advance\ps@tempdima \fboxsep \advance\ps@tempdima \dp\ps@tempboxa
   \hbox{\lower \ps@tempdima\hbox
  {\vbox{\hrule height \fboxrule
          \hbox{\vrule width \fboxrule \hskip\fboxsep
          \vbox{\vskip\fboxsep \box\ps@tempboxa\vskip\fboxsep}\hskip 
                 \fboxsep\vrule width \fboxrule}
                 \hrule height \fboxrule}}}}
\fi
\newread\ps@stream
\newif\ifnot@eof       
\newif\if@noisy        
\newif\if@atend        
\newif\if@psfile       
{\catcode`\%=12\global\gdef\epsf@start{
\def\epsf@PS{PS}
\def\epsf@getbb#1{%
\openin\ps@stream=#1
\ifeof\ps@stream\ps@typeout{Error, File #1 not found}\else
   {\not@eoftrue \chardef\other=12
    \def\do##1{\catcode`##1=\other}\dospecials \catcode`\ =10
    \loop
       \if@psfile
	  \read\ps@stream to \epsf@fileline
       \else{
	  \obeyspaces
          \read\ps@stream to \epsf@tmp\global\let\epsf@fileline\epsf@tmp}
       \fi
       \ifeof\ps@stream\not@eoffalse\else
       \if@psfile\else
       \expandafter\epsf@test\epsf@fileline:. \\%
       \fi
          \expandafter\epsf@aux\epsf@fileline:. \\%
       \fi
   \ifnot@eof\repeat
   }\closein\ps@stream\fi}%
\long\def\epsf@test#1#2#3:#4\\{\def\epsf@testit{#1#2}
			\ifx\epsf@testit\epsf@start\else
\ps@typeout{Warning! File does not start with `\epsf@start'.  It may not be a PostScript file.}
			\fi
			\@psfiletrue} 
{\catcode`\%=12\global\let\epsf@percent=
\long\def\epsf@aux#1#2:#3\\{\ifx#1\epsf@percent
   \def\epsf@testit{#2}\ifx\epsf@testit\epsf@bblit
	\@atendfalse
        \epsf@atend #3 . \\%
	\if@atend	
	   \if@verbose{
		\ps@typeout{psfig: found `(atend)'; continuing search}
	   }\fi
        \else
        \epsf@grab #3 . . . \\%
        \not@eoffalse
        \global\no@bbfalse
        \fi
   \fi\fi}%
\def\epsf@grab #1 #2 #3 #4 #5\\{%
   \global\def\epsf@llx{#1}\ifx\epsf@llx\empty
      \epsf@grab #2 #3 #4 #5 .\\\else
   \global\def\epsf@lly{#2}%
   \global\def\epsf@urx{#3}\global\def\epsf@ury{#4}\fi}%
\def\epsf@atendlit{(atend)} 
\def\epsf@atend #1 #2 #3\\{%
   \def\epsf@tmp{#1}\ifx\epsf@tmp\empty
      \epsf@atend #2 #3 .\\\else
   \ifx\epsf@tmp\epsf@atendlit\@atendtrue\fi\fi}

\chardef\psletter = 11 
\chardef\other = 12

\newif \ifdebug 
\newif\ifc@mpute 
\c@mputetrue 

\let\then = \relax
\def\r@dian{pt }
\let\r@dians = \r@dian
\let\dimensionless@nit = \r@dian
\let\dimensionless@nits = \dimensionless@nit
\def\internal@nit{sp }
\let\internal@nits = \internal@nit
\newif\ifstillc@nverging
\def \Mess@ge #1{\ifdebug \then \message {#1} \fi}

{ 
	\catcode `\@ = \psletter
	\gdef \nodimen {\expandafter \n@dimen \the \dimen}
	\gdef \term #1 #2 #3%
	       {\edef \t@ {\the #1}
		\edef \t@@ {\expandafter \n@dimen \the #2\r@dian}%
		\t@rm {\t@} {\t@@} {#3}%
	       }
	\gdef \t@rm #1 #2 #3%
	       {{%
		\count 0 = 0
		\dimen 0 = 1 \dimensionless@nit
		\dimen 2 = #2\relax
		\Mess@ge {Calculating term #1 of \nodimen 2}%
		\loop
		\ifnum	\count 0 < #1
		\then	\advance \count 0 by 1
			\Mess@ge {Iteration \the \count 0 \space}%
			\Multiply \dimen 0 by {\dimen 2}%
			\Mess@ge {After multiplication, term = \nodimen 0}%
			\Divide \dimen 0 by {\count 0}%
			\Mess@ge {After division, term = \nodimen 0}%
		\repeat
		\Mess@ge {Final value for term #1 of 
				\nodimen 2 \space is \nodimen 0}%
		\xdef \Term {#3 = \nodimen 0 \r@dians}%
		\aftergroup \Term
	       }}
	\catcode `\p = \other
	\catcode `\t = \other
	\gdef \n@dimen #1pt{#1} 
}

\def \Divide #1by #2{\divide #1 by #2} 

\def \Multiply #1by #2
       {{
	\count 0 = #1\relax
	\count 2 = #2\relax
	\count 4 = 65536
	\Mess@ge {Before scaling, count 0 = \the \count 0 \space and
			count 2 = \the \count 2}%
	\ifnum	\count 0 > 32767 
	\then	\divide \count 0 by 4
		\divide \count 4 by 4
	\else	\ifnum	\count 0 < -32767
		\then	\divide \count 0 by 4
			\divide \count 4 by 4
		\else
		\fi
	\fi
	\ifnum	\count 2 > 32767 
	\then	\divide \count 2 by 4
		\divide \count 4 by 4
	\else	\ifnum	\count 2 < -32767
		\then	\divide \count 2 by 4
			\divide \count 4 by 4
		\else
		\fi
	\fi
	\multiply \count 0 by \count 2
	\divide \count 0 by \count 4
	\xdef \product {#1 = \the \count 0 \internal@nits}%
	\aftergroup \product
       }}

\def\r@duce{\ifdim\dimen0 > 90\r@dian \then   
		\multiply\dimen0 by -1
		\advance\dimen0 by 180\r@dian
		\r@duce
	    \else \ifdim\dimen0 < -90\r@dian \then  
		\advance\dimen0 by 360\r@dian
		\r@duce
		\fi
	    \fi}

\def\Sine#1%
       {{%
	\dimen 0 = #1 \r@dian
	\r@duce
	\ifdim\dimen0 = -90\r@dian \then
	   \dimen4 = -1\r@dian
	   \c@mputefalse
	\fi
	\ifdim\dimen0 = 90\r@dian \then
	   \dimen4 = 1\r@dian
	   \c@mputefalse
	\fi
	\ifdim\dimen0 = 0\r@dian \then
	   \dimen4 = 0\r@dian
	   \c@mputefalse
	\fi
	\ifc@mpute \then
		\divide\dimen0 by 180
		\dimen0=3.141592654\dimen0
		\dimen 2 = 3.1415926535897963\r@dian 
		\divide\dimen 2 by 2 
		\Mess@ge {Sin: calculating Sin of \nodimen 0}%
		\count 0 = 1 
		\dimen 2 = 1 \r@dian 
		\dimen 4 = 0 \r@dian 
		\loop
			\ifnum	\dimen 2 = 0 
			\then	\stillc@nvergingfalse 
			\else	\stillc@nvergingtrue
			\fi
			\ifstillc@nverging 
			\then	\term {\count 0} {\dimen 0} {\dimen 2}%
				\advance \count 0 by 2
				\count 2 = \count 0
				\divide \count 2 by 2
				\ifodd	\count 2 
				\then	\advance \dimen 4 by \dimen 2
				\else	\advance \dimen 4 by -\dimen 2
				\fi
		\repeat
	\fi		
			\xdef \sine {\nodimen 4}%
       }}

\def\Cosine#1{\ifx\sine\UnDefined\edef\Savesine{\relax}\else
		             \edef\Savesine{\sine}\fi
	{\dimen0=#1\r@dian\advance\dimen0 by 90\r@dian
	 \Sine{\nodimen 0}
	 \xdef\cosine{\sine}
	 \xdef\sine{\Savesine}}}	      

\def\psdraft{
	\def\@psdraft{0}
}
\def\psfull{
	\def\@psdraft{100}
}

\psfull

\newif\if@scalefirst
\def\psscalefirst{\@scalefirsttrue}
\def\psrotatefirst{\@scalefirstfalse}
\psrotatefirst

\newif\if@draftbox
\def\psnodraftbox{
	\@draftboxfalse
}
\def\psdraftbox{
	\@draftboxtrue
}
\@draftboxtrue

\newif\if@prologfile
\newif\if@postlogfile
\def\pssilent{
	\@noisyfalse
}
\def\psnoisy{
	\@noisytrue
}
\psnoisy
\newif\if@bbllx
\newif\if@bblly
\newif\if@bburx
\newif\if@bbury
\newif\if@height
\newif\if@width
\newif\if@rheight
\newif\if@rwidth
\newif\if@angle
\newif\if@clip
\newif\if@verbose
\newif\if@scale
\def\@p@@sclip#1{\@cliptrue}

\newif\if@decmpr

\def\@p@@sfigure#1{\def\@p@sfile{null}\def\@p@sbbfile{null}
	        \openin1=#1.bb
		\ifeof1\closein1
	        	\openin1=\figurepath#1.bb
			\ifeof1\closein1
			        \openin1=#1
				\ifeof1\closein1%
				       \openin1=\figurepath#1
					\ifeof1
					   \ps@typeout{Error, File #1 not found}
						\if@bbllx\if@bblly
				   		\if@bburx\if@bbury
			      				\def\@p@sfile{#1}%
			      				\def\@p@sbbfile{#1}%
							\@decmprfalse
				  	   	\fi\fi\fi\fi
					\else\closein1
				    		\def\@p@sfile{\figurepath#1}%
				    		\def\@p@sbbfile{\figurepath#1}%
						\@decmprfalse
	                       		\fi%
			 	\else\closein1%
					\def\@p@sfile{#1}
					\def\@p@sbbfile{#1}
					\@decmprfalse
			 	\fi
			\else
				\def\@p@sfile{\figurepath#1}
				\def\@p@sbbfile{\figurepath#1.bb}
				\@decmprtrue
			\fi
		\else
			\def\@p@sfile{#1}
			\def\@p@sbbfile{#1.bb}
			\@decmprtrue
		\fi}

\def\@p@@sfile#1{\@p@@sfigure{#1}}

\def\@p@@sbbllx#1{
		\@bbllxtrue
		\dimen100=#1
		\edef\@p@sbbllx{\number\dimen100}
}
\def\@p@@sbblly#1{
		\@bbllytrue
		\dimen100=#1
		\edef\@p@sbblly{\number\dimen100}
}
\def\@p@@sbburx#1{
		\@bburxtrue
		\dimen100=#1
		\edef\@p@sbburx{\number\dimen100}
}
\def\@p@@sbbury#1{
		\@bburytrue
		\dimen100=#1
		\edef\@p@sbbury{\number\dimen100}
}
\def\@p@@sheight#1{
		\@heighttrue
		\dimen100=#1
   		\edef\@p@sheight{\number\dimen100}
}
\def\@p@@swidth#1{
		\@widthtrue
		\dimen100=#1
		\edef\@p@swidth{\number\dimen100}
}
\def\@p@@srheight#1{
		\@rheighttrue
		\dimen100=#1
		\edef\@p@srheight{\number\dimen100}
}
\def\@p@@srwidth#1{
		\@rwidthtrue
		\dimen100=#1
		\edef\@p@srwidth{\number\dimen100}
}
\def\@p@@sangle#1{
		\@angletrue
		\edef\@p@sangle{#1} 
}
\def\@p@@srotate#1{\@p@@sangle{-#1}}
\def\@p@@sscale#1{
		\@scaletrue
		\edef\@p@sscale{#1}
}
\def\@p@@ssilent#1{ 
		\@verbosefalse
}
\def\@p@@sprolog#1{\@prologfiletrue\def\@prologfileval{#1}}
\def\@p@@spostlog#1{\@postlogfiletrue\def\@postlogfileval{#1}}
\def\@cs@name#1{\csname #1\endcsname}
\def\@setparms#1=#2,{\@cs@name{@p@@s#1}{#2}}
\def\ps@init@parms{
		\@bbllxfalse \@bbllyfalse
		\@bburxfalse \@bburyfalse
		\@heightfalse \@widthfalse
		\@rheightfalse \@rwidthfalse
		\@scalefalse
		\def\@p@sbbllx{}\def\@p@sbblly{}
		\def\@p@sbburx{}\def\@p@sbbury{}
		\def\@p@sheight{}\def\@p@swidth{}
		\def\@p@srheight{}\def\@p@srwidth{}
		\def\@p@sangle{0}
		\def\@p@sfile{} \def\@p@sbbfile{}
		\def\@p@scost{10}
		\def\@sc{}
		\@prologfilefalse
		\@postlogfilefalse
		\@clipfalse
		\if@noisy
			\@verbosetrue
		\else
			\@verbosefalse
		\fi
}
\def\parse@ps@parms#1{
	 	\@psdo\@psfiga:=#1\do
		   {\expandafter\@setparms\@psfiga,}}
\newif\ifno@bb
\def\bb@missing{
	\if@verbose{
		\ps@typeout{psfig: searching \@p@sbbfile \space  for bounding box}
	}\fi
	\no@bbtrue
	\epsf@getbb{\@p@sbbfile}
        \ifno@bb \else \bb@cull\epsf@llx\epsf@lly\epsf@urx\epsf@ury\fi
}	
\def\bb@cull#1#2#3#4{
	\dimen100=#1 bp\edef\@p@sbbllx{\number\dimen100}
	\dimen100=#2 bp\edef\@p@sbblly{\number\dimen100}
	\dimen100=#3 bp\edef\@p@sbburx{\number\dimen100}
	\dimen100=#4 bp\edef\@p@sbbury{\number\dimen100}
	\no@bbfalse
}
\newdimen\p@intvaluex
\newdimen\p@intvaluey
\def\rotate@#1#2{{\dimen0=#1 sp\dimen1=#2 sp
		  \global\p@intvaluex=\cosine\dimen0
		  \dimen3=\sine\dimen1
		  \global\advance\p@intvaluex by -\dimen3
		  \global\p@intvaluey=\sine\dimen0
		  \dimen3=\cosine\dimen1
		  \global\advance\p@intvaluey by \dimen3
		  }}
\def\compute@bb{
		\no@bbfalse
		\if@bbllx \else \no@bbtrue \fi
		\if@bblly \else \no@bbtrue \fi
		\if@bburx \else \no@bbtrue \fi
		\if@bbury \else \no@bbtrue \fi
		\ifno@bb \bb@missing \fi
		\ifno@bb \ps@typeout{FATAL ERROR: no bb supplied or found}
			\no-bb-error
		\fi
		\count203=\@p@sbburx
		\count204=\@p@sbbury
		\advance\count203 by -\@p@sbbllx
		\advance\count204 by -\@p@sbblly
		\edef\ps@bbw{\number\count203}
		\edef\ps@bbh{\number\count204}
		\if@angle 
			\Sine{\@p@sangle}\Cosine{\@p@sangle}
	        	{\dimen100=\maxdimen\xdef\r@p@sbbllx{\number\dimen100}
					    \xdef\r@p@sbblly{\number\dimen100}
			                    \xdef\r@p@sbburx{-\number\dimen100}
					    \xdef\r@p@sbbury{-\number\dimen100}}
                        \def\minmaxtest{
			   \ifnum\number\p@intvaluex<\r@p@sbbllx
			      \xdef\r@p@sbbllx{\number\p@intvaluex}\fi
			   \ifnum\number\p@intvaluex>\r@p@sbburx
			      \xdef\r@p@sbburx{\number\p@intvaluex}\fi
			   \ifnum\number\p@intvaluey<\r@p@sbblly
			      \xdef\r@p@sbblly{\number\p@intvaluey}\fi
			   \ifnum\number\p@intvaluey>\r@p@sbbury
			      \xdef\r@p@sbbury{\number\p@intvaluey}\fi
			   }
			\rotate@{\@p@sbbllx}{\@p@sbblly}
			\minmaxtest
			\rotate@{\@p@sbbllx}{\@p@sbbury}
			\minmaxtest
			\rotate@{\@p@sbburx}{\@p@sbblly}
			\minmaxtest
			\rotate@{\@p@sbburx}{\@p@sbbury}
			\minmaxtest
			\edef\@p@sbbllx{\r@p@sbbllx}\edef\@p@sbblly{\r@p@sbblly}
			\edef\@p@sbburx{\r@p@sbburx}\edef\@p@sbbury{\r@p@sbbury}
		\fi
		\count203=\@p@sbburx
		\count204=\@p@sbbury
		\advance\count203 by -\@p@sbbllx
		\advance\count204 by -\@p@sbblly
		\edef\@bbw{\number\count203}
		\edef\@bbh{\number\count204}
}
\def\in@hundreds#1#2#3{\count240=#2 \count241=#3
		     \count100=\count240	
		     \divide\count100 by \count241
		     \count101=\count100
		     \multiply\count101 by \count241
		     \advance\count240 by -\count101
		     \multiply\count240 by 10
		     \count101=\count240	
		     \divide\count101 by \count241
		     \count102=\count101
		     \multiply\count102 by \count241
		     \advance\count240 by -\count102
		     \multiply\count240 by 10
		     \count102=\count240	
		     \divide\count102 by \count241
		     \count200=#1\count205=0
		     \count201=\count200
			\multiply\count201 by \count100
		 	\advance\count205 by \count201
		     \count201=\count200
			\divide\count201 by 10
			\multiply\count201 by \count101
			\advance\count205 by \count201
		     \count201=\count200
			\divide\count201 by 100
			\multiply\count201 by \count102
			\advance\count205 by \count201
		     \edef\@result{\number\count205}
}
\def\ps@scaleinhundreds#1{
		\in@hundreds{#1}{\@p@sscale}{100}
		\edef#1{\@result}
}
\def\compute@wfromh{
		\in@hundreds{\@p@sheight}{\@bbw}{\@bbh}
		\edef\@p@swidth{\@result}
}
\def\compute@hfromw{
	        \in@hundreds{\@p@swidth}{\@bbh}{\@bbw}
		\edef\@p@sheight{\@result}
}
\def\compute@handw{
		\if@height 
			\if@width
			\else
				\compute@wfromh
			\fi
		\else 
			\if@width
				\compute@hfromw
			\else
				\edef\@p@sheight{\@bbh}
				\edef\@p@swidth{\@bbw}
			\fi
		\fi
}
\def\compute@resv{
		\if@rheight \else \edef\@p@srheight{\@p@sheight} \fi
		\if@rwidth \else \edef\@p@srwidth{\@p@swidth} \fi
}
\def\compute@sizes{
	\compute@bb
	\if@scalefirst\if@angle
	\if@width
	   \in@hundreds{\@p@swidth}{\@bbw}{\ps@bbw}
	   \edef\@p@swidth{\@result}
	\fi
	\if@height
	   \in@hundreds{\@p@sheight}{\@bbh}{\ps@bbh}
	   \edef\@p@sheight{\@result}
	\fi
	\fi\fi
	\compute@handw
	\compute@resv
	\if@scale
	   \if@verbose
	      \ps@typeout{(scaling by \@p@sscale)}%
	   \fi
	   \ps@scaleinhundreds{\@p@swidth}%
	   \ps@scaleinhundreds{\@p@sheight}%
	   \ps@scaleinhundreds{\@p@srwidth}%
	   \ps@scaleinhundreds{\@p@srheight}%
	\fi
}

\def\psfig#1{\vbox {
	\ps@init@parms
	\parse@ps@parms{#1}
	\compute@sizes
	\ifnum\@p@scost<\@psdraft{
		\special{ps::[begin] 	\@p@swidth \space \@p@sheight \space
				\@p@sbbllx \space \@p@sbblly \space
				\@p@sbburx \space \@p@sbbury \space
				startTexFig \space }
		\if@angle
			\special {ps:: \@p@sangle \space rotate \space} 
		\fi
		\if@clip{
			\if@verbose{
				\ps@typeout{(clip)}
			}\fi
			\special{ps:: doclip \space }
		}\fi
		\if@prologfile
		    \special{ps: plotfile \@prologfileval \space } \fi
		\if@decmpr{
			\if@verbose{
				\ps@typeout{psfig: including \@p@sfile.Z \space }
			}\fi
			\special{ps: plotfile "`zcat \@p@sfile.Z" \space }
		}\else{
			\if@verbose{
				\ps@typeout{psfig: including \@p@sfile \space }
			}\fi
			\special{ps: plotfile \@p@sfile \space }
		}\fi
		\if@postlogfile
		    \special{ps: plotfile \@postlogfileval \space } \fi
		\special{ps::[end] endTexFig \space }
		\vbox to \@p@srheight true sp{
			\hbox to \@p@srwidth true sp{
				\hss
			}
		\vss
		}
	}\else{
			\vbox to \@p@srheight true sp{
			\vss
			\hbox to \@p@srwidth true sp{\hss}
			\vss
			}

	}\fi
}}
\psfigRestoreAt
\let\@=\LaTeXAtSign

\newcommand{\beq}{\begin{equation}}
\newcommand{\eeq}{\end{equation}}
\newcommand{\beqnn}{\begin{displaymath}}	
\newcommand{\eeqnn}{\end{displaymath}}		
\newcommand{\beqa}{\begin{eqnarray}}
\newcommand{\eeqa}{\end{eqnarray}}
\newcommand{\beqann}{\begin{eqnarray*}}
\newcommand{\eeqann}{\end{eqnarray*}}
\newcommand{\nn}{\nonumber}
\newcommand{\ben}{\begin{enumerate}}
\newcommand{\een}{\end{enumerate}}
\newcommand{\bit}{\begin{itemize}}
\newcommand{\eit}{\end{itemize}}
\newcommand{\bc}{\begin{center}}
\newcommand{\ec}{\end{center}}

\newcommand{\tabref}[1]{Table~\ref{#1}}
\newcommand{\tabsref}[2]{Tables~\ref{#1} and~\ref{#2}}
\newcommand{\figref}[1]{Figure~\ref{#1}}
\newcommand{\figrefbare}[1]{\ref{#1}}
\newcommand{\figrefs}[1]{Figures~\ref{#1}}
\newcommand{\figsref}[2]{Figures~\ref{#1} and~\ref{#2}}
\newcommand{\eqref}[1]{Equation~\ref{#1}}
\newcommand{\eqsref}[2]{Equations~\ref{#1} and~\ref{#2}}
\newcommand{\chapref}[1]{Chapter~\ref{#1}}
\newcommand{\secref}[1]{Section~\ref{#1}}
\newcommand{\secsref}[2]{Sections~\ref{#1} and~\ref{#2}}
\newcommand{\appref}[1]{Appendix~\ref{#1}}

\newcommand{\Tabref}[1]{\tabref{#1}}
\newcommand{\Tabsref}[2]{\tabsref{#1}{#2}}
\newcommand{\Figref}[1]{\figref{#1}}
\newcommand{\Figrefs}[1]{\figrefs{#1}}
\newcommand{\Figsref}[2]{\figsref{#1}{#2}}
\newcommand{\Eqref}[1]{\eqref{#1}}
\newcommand{\Eqsref}[2]{\eqsref{#1}{#2}}
\newcommand{\Chapref}[1]{\chapref{#1}}
\newcommand{\Secref}[1]{\secref{#1}}
\newcommand{\Appref}[1]{\appref{#1}}

\newcommand{\equals}[1]{\mbox{$\stackrel{\rm #1}{=}$}}

\newcommand{\coude}{coud\'{e}}
\newcommand{\ie}{i.e.,}
\newcommand{\eg}{e.g.,}
\newcommand{\AIPS}{\mbox{$\cal AIPS$}}

\newcommand{\Htwo}{\mbox{H\,{\sc ii}}}

\newcommand{\eBoo}{\mbox{$\eta$~Boo}}
\newcommand{\aCen}{\mbox{$\alpha$~Cen}}
\newcommand{\bCen}{\mbox{$\beta$~Cen}}
\newcommand{\eCar}{\mbox{$\eta$~Car}}
\newcommand{\oCet}{\mbox{$o$~Cet}}
\newcommand{\gCru}{\mbox{$\gamma$~Cru}}
\newcommand{\eEri}{\mbox{$\eta$~Eri}}
\newcommand{\aGru}{\mbox{$\alpha$~Gru}}
\newcommand{\bGru}{\mbox{$\beta$~Gru}}
\newcommand{\aHer}{\mbox{$\alpha$~Her}}
\newcommand{\iLib}{\mbox{$\iota^{\scriptscriptstyle 1}$~Lib}}
\newcommand{\nOct}{\mbox{$\nu$~Oct}}
\newcommand{\eOph}{\mbox{$\eta$~Oph}}
\newcommand{\aOri}{\mbox{$\alpha$~Ori}}
\newcommand{\gOri}{\mbox{$\gamma$~Ori}}
\newcommand{\aPav}{\mbox{$\alpha$~Pav}}
\newcommand{\gRet}{\mbox{$\gamma$~Ret}}
\newcommand{\aSco}{\mbox{$\alpha$~Sco}}
\newcommand{\eSco}{\mbox{$\varepsilon$~Sco}}
\newcommand{\tSco}{\mbox{$\tau$~Sco}}
\newcommand{\dSco}{\mbox{$\delta$~Sco}}
\newcommand{\lSgr}{\mbox{$\lambda$~Sgr}}
\newcommand{\sSgr}{\mbox{$\sigma$~Sgr}}
\newcommand{\aTau}{\mbox{$\alpha$~Tau}}
\newcommand{\aTuc}{\mbox{$\alpha$~Tuc}}

\newcommand{\about}{\mbox{$\sim$\,}}	
\newcommand{\degree}{\mbox{$^\circ$}}
\newcommand{\h}{\mbox{$^{\rm h}$}}
\newcommand{\m}{\mbox{$^{\rm m}$}}

\let\mc=\multicolumn
\let\ph=\phantom
\newcommand{\phz}{\phantom{0}}
\newcommand{\Q}{\rlap{:}}
\newcommand{\mcc}[1]{\mc{1}{c}{#1}}
\newcommand{\mcl}[1]{\mc{1}{l}{#1}}
\newcommand{\mcr}[1]{\mc{1}{r}{#1}}

\def\Msol{\mbox{${M}_\odot$}}
\def\Lsol{\mbox{${L}_\odot$}}
\def\Rsol{\mbox{${R}_\odot$}}
\def\gsol{\mbox{${g}_\odot$}}

\def\deg{\hbox{$^\circ$}}
\def\solar{\mbox{$_{\odot}$}}
\def\sun{\hbox{$\odot$}}
\def\earth{\hbox{$\oplus$}}
\def\la{\mathrel{\hbox{\rlap{\hbox{\lower4pt\hbox{$\sim$}}}\hbox{$<$}}}}
\def\ga{\mathrel{\hbox{\rlap{\hbox{\lower4pt\hbox{$\sim$}}}\hbox{$>$}}}}
\def\sq{\hbox{\rlap{$\sqcap$}$\sqcup$}}
\def\arcmin{\hbox{$^\prime$}}
\def\arcsec{\hbox{$^{\prime\prime}$}}
\def\fd{\hbox{$.\!\!^{\rm d}$}}
\def\fh{\hbox{$.\!\!^{\rm h}$}}
\def\fm{\hbox{$.\!\!^{\rm m}$}}
\def\fs{\hbox{$.\!\!^{\rm s}$}}
\def\fdg{\hbox{$.\!\!^\circ$}}
\def\farcm{\hbox{$.\mkern-4mu^\prime$}}
\def\farcs{\hbox{$.\!\!^{\prime\prime}$}}
\def\fp{\hbox{$.\!\!^{\scriptscriptstyle\rm p}$}}
\def\micron{\hbox{$\mu$m}}

\newcommand{\rz}{\mbox{$r_0$}}
\newcommand{\tz}{\mbox{$t_0$}}

\newcommand{\half}{{\textstyle\frac{1}{2}}}
\newcommand{\threehalves}{{\textstyle\frac{3}{2}}}
\newcommand{\quarter}{{\textstyle\frac{1}{4}}}
\newcommand{\third}{{\textstyle\frac{1}{3}}}

\def\laeq{\lower.5ex\hbox{{$\:\scriptstyle\buildrel < \over \sim\:$}}}
\def\gaeq{\lower.5ex\hbox{{$\:\scriptstyle\buildrel > \over \sim\:$}}}

\makeatletter 
 \def\sub#1{\relax\ifmmode _{\fam\z@ #1}\else
         $_{\fam\z@ #1}$\fi}
 \def\super#1{\relax\ifmmode ^{\fam\z@ #1}\else
         $^{\fam\z@ #1}$\fi}
\makeatother 

\newcommand{\downup}[3]{#1\sub{\rm #2}\super{\rm #3}}
\newcommand{\down}[2]{#1\sub{#2}}
\newcommand{\up}[2]{#1\super{#2}}

\newcommand{\mycaption}[3]
{\if*#2 \caption{#3\label{#1}}
 \else  \caption[#2]{#3\label{#1}}
 \fi}

\newcommand{\comment}[1]{\relax}

\long\def\COMMENT#1\ENDCOMMENT{}
\def\ENDCOMMENT{}

\newcommand{\Aosc}{\down{A}{osc}}
\newcommand{\DLbol}{\down{\DL}{bol}}
\newcommand{\DLlambda}{\DL_{\lambda}}
\newcommand{\DL}{(\delta L/L)}
\newcommand{\DW}{(\delta W/W)}
\newcommand{\DT}{\overline{\Delta T}}
\newcommand{\dnuz}{\delta \nu_0}
\newcommand{\Dnuz}{\Delta \nu_0}
\newcommand{\Dnu}[1]{\Delta \nu_{#1}}
\newcommand{\dnu}[1]{\delta \nu_{#1}}
\newcommand{\DvSun}{\Dv\commaSun}
\newcommand{\Dv}{\down{v}{osc}}
\newcommand{\Fcon}{\down{F}{con}}
\newcommand{\HP}{\down{H}{P}}
\newcommand{\Hrho}{H_{\rho}}
\newcommand{\cP}{\down{c}{P}}
\newcommand{\cms}{\mbox{cm\,s$^{-1}$}}
\newcommand{\commaSun}{\mbox{$_{,\odot}$}}
\newcommand{\cs}{\down{c}{s}}
\newcommand{\kms}{\mbox{km\,s$^{-1}$}}
\newcommand{\lambdabol}{\down{\lambda}{bol}}
\newcommand{\mean}[1]{\langle#1\rangle}
\newcommand{\ms}{\mbox{m\,s$^{-1}$}}
\newcommand{\muHz}{\mbox{$\mu$Hz}}
\newcommand{\nmax}{\down{n}{max}}
\newcommand{\nuac}{\down{\nu}{ac}}
\newcommand{\numax}{\down{\nu}{max}}
\newcommand{\nurot}{\down{\nu}{rot}}
\newcommand{\sigmaRMS}{\down{\sigma}{rms}}
\newcommand{\sigmaAMP}{\down{\sigma}{amp}}
\newcommand{\sigmaPS}{\down{\sigma}{PS}}
\newcommand{\tsub}{\down{t}{sub}}
\newcommand{\vcon}{\down{v}{con}}

\newcommand{\epsEri}{\mbox{$\varepsilon$~Eri}}
\newcommand{\aCenA}{\mbox{$\alpha$~Cen~A}}
\newcommand{\bHyi}{\mbox{$\beta$~Hyi}}

\newcommand{\PSPS}{PS$\otimes$PS}
\newcommand{\Teff}{\down{T}{eff}}

\ifnfsstwo
  \newcommand{\mitbf}[1] {\hbox{\mathversion{bold}$#1$}}
  \newcommand{\rmn}[1] {{\mathrm #1}}
  \newcommand{\itl}[1] {{\mathit #1}}
  \newcommand{\bld}[1] {{\mathbf #1}}
\fi

\ifnfssone
  \newmathalphabet{\mathit}
    \addtoversion{normal}{\mathit}{cmr}{m}{it}
    \addtoversion{bold}{\mathit}{cmr}{bx}{it}
    \newcommand{\mitbf}[1] {\hbox{\mathversion{bold}$#1$}}
    \newcommand{\rmn}[1] {{\mathrm #1}}
    \newcommand{\itl}[1] {{\mathit #1}}
    \newcommand{\bld}[1] {{\mathbf #1}}
\fi

\ifoldfss    
  \newcommand{\mitbf}[1] {\mbox{\boldmath $#1$}}
  \newcommand{\rmn}[1] {{\rm #1}}
  \newcommand{\itl}[1] {{\it #1}}
  \newcommand{\bld}[1] {{\bf #1}}
\fi

\loadboldmathitalic
\loadboldgreek

\title[Angular diameter of R~Dor]
{The angular diameter of R~Doradus: a nearby Mira-like star}
\author[T.~R.~Bedding et al.]
       {T.~R.~Bedding,$^1$\thanks{E-mail: \tt bedding@physics.usyd.edu.au}
	A.~A.~Zijlstra,$^2$ O.~von~der~L\"uhe,$^2$ 
	J.~G.~Robertson,$^1$ \cr 
	R.~G.~Marson,$^{1,3}$
	J.~R.~Barton$^4$
	and B.~S.~Carter$^{5,6}$\\
	$^1$School of Physics, University of Sydney 2006, Australia\\
	$^2$European Southern Observatory, Karl-Schwarzschild-Str.~2,
		D-85748 Garching bei M\"unchen, Germany\\
	$^3$Current address: NRAO Array Operations Center, P.O. Box 0,
	Socorro NM 87801, USA\\
	$^4$Anglo-Australian Observatory, P.O. Box 296, Epping 2121,
	Australia\\
	$^5$South African Astronomical Observatory, P.O. Box 9, Observatory
	7935, South Africa\\
	$^6$Current address: Carter Observatory, P.O. Box 2909, Wellington,
	New Zealand
}
\pubyear{1996}

\begin{document}

\maketitle
\begin{abstract} We find the angular diameter of R~Doradus to be
$57\pm5$\,mas, exceeding that of Betelgeuse and implying that R~Dor is
larger in apparent size than every star except the Sun.  R~Dor is shown to
be closely related to the Mira variables.  We estimate an effective
temperature of $2740\pm190$\,K, a distance of $61\pm7$\,pc, a luminosity of
$6500\pm1400$\,\Lsol\ and a radius of $370\pm50$\,\Rsol.  The
characteristics of R Dor are consistent with it being near the edge of a
Mira instability strip.  We detect non-zero closure phases from R~Dor,
indicating an asymmetric brightness distribution.  We also observed W~Hya,
a small-amplitude Mira, for which we find an angular diameter of
$44\pm4$\,mas.
\end{abstract}


\section{Introduction}

Several nearby red giant and supergiant stars have been resolved with 4-m
telescopes, producing a number of significant results.  Angular diameter
measurements have provided evidence for first-overtone pulsation in Mira
variables \cite{THB94b,HST95} and produced a new $\log\Teff$ versus ($J-K$)
relation \cite{Fea96}.  Also important are the discoveries of hotspots on
\aOri\ (Betelgeuse) by \cite{BHB90} and of asymmetries in the atmosphere of
\oCet\ and similar stars \cite{WBB92,HGG92,THB94a}.

It has long been assumed that \aOri, a red supergiant, has the largest
angular diameter of any star in the night sky.  Here we report measurements
of the red giant R~Doradus (HR~1492), a semiregular variable of spectral
type M8, which show it to exceed \aOri\ in angular size.  Although the $V$
magnitude of R~Dor is only 5.4, at infrared wavelengths it rivals \aOri\ as
the brightest star in the sky.  This has lead \citeone{Win71} to predict a
large angular size but R~Dor has not previously been observed at high
angular resolution, presumably due to its southerly declination.  We also
report observations of W~Hydrae (HD~120285), which is a small-amplitude
Mira variable that also has a large angular diameter.

\section{Observations}

We have carried out observations of R~Dor in the near infrared
(1.25\,\micron) and the red (855\,nm) using aperture-masking.  This
technique involves modifying the telescope pupil with a mask, the rationale
for which has been discussed elsewhere \cite{H+B92,B+H93,BvdLZ93,Han94}.
In brief, an aperture mask with a small number of non-redundantly spaced
holes provides more accurately calibrated measurements at the full
diffraction limit.  This comes at the expense of lower sensitivity and
poorer spatial-frequency coverage, and a good compromise in the photon-rich
infrared regime is to use an annular mask \cite{H+B92}.  This has full
spatial frequency coverage while being minimally redundant: roughly
speaking, each baseline is measured twice.  Furthermore, a thin annulus
whose width is comparable to the atmospheric coherence length ($r_0$)
largely retains the advantages of accurate calibration and enhanced
resolution.  To date, however, results from annular masks have only been
published for binary stars \cite{HBC89}.

In the visible regime, where photon noise generally dominates over
atmospheric noise, an annular mask is not a good choice.  Instead, improved
sensitivity and spatial-frequency coverage is best achieved by using a long
slit \cite{A+R77,B+H93}.

\subsection{NTT Observations}

Observations with the 3.5-m New Technology Telescope (NTT) were made with
the SHARP infrared camera \cite{HBD92}.  Details of the experimental setup
are given in \citetwo{BvdLZ93}{BvdLZ95}.  We observed R~Dor on 1993
August~7 using an aperture mask with seven holes arranged in a
non-redundant two-dimensional configuration.  Each hole had an effective
diameter (as projected onto the primary mirror) of 25\,cm and lay on a
circle of diameter 3.05\,m.  The wavelength was 1.25\,\micron, selected
using a standard $J$ filter.  For both R~Dor and a calibrator star (\gRet),
we obtained 500 short (0.1\,s) exposures.

Fringe visibilities were extracted from the averaged power spectrum and
were calibrated for atmospheric and instrumental effects by dividing by
visibilities from the calibrator star.  The upper panel of
\figref{fig.vis-NTT1993} shows these calibrated visibilities as a function
of baseline length.  It is clear that R~Doradus is resolved, despite the
fact that the visibility at this wavelength does not fall below 50\%, even
on the longest baselines.  There is a scatter in the visibility
measurements at fixed baselines whose cause is unclear, although it may be
due to a mismatch in seeing conditions for observations of R~Dor and its
calibrator star.  Fitting a uniform-disk model to these data gives a
diameter of $(57\pm5)$\,mas (milliarcsec).

\if\preprint1
	\begin{figure}
\centerline{
\psfig{figure=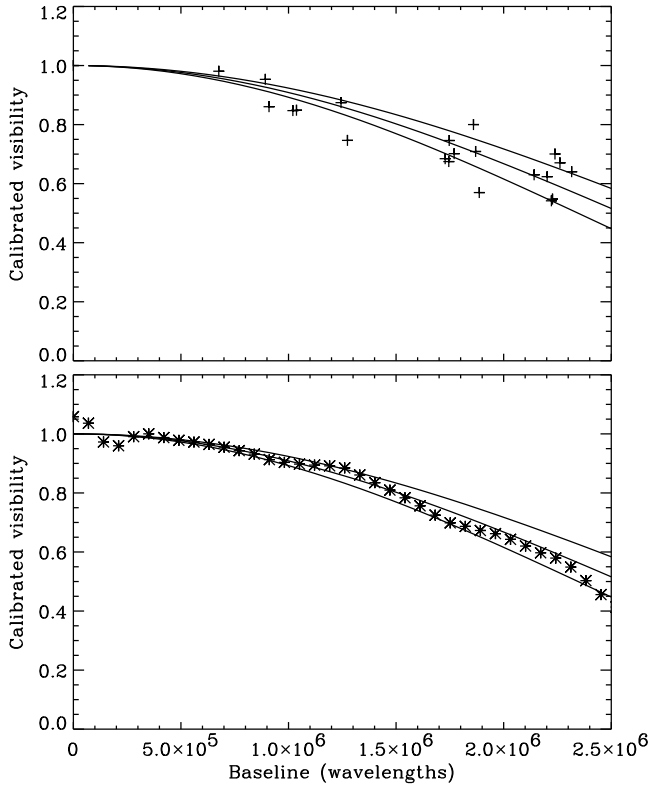%
,bbllx=93pt,bblly=207pt,bburx=517pt,bbury=726pt%
,width=\the\hsize,scale=100}}

\caption[]{\label{fig.vis-NTT1993} Visibilities of R~Dor measured at
1.25\,\micron\ with the NTT in 1993 August.  The upper panel is for the
7-hole mask, with crosses showing visibilities on 21 baselines.  The lower
panel is for the annular mask, with data points being azimuthal averages.
In both panels the solid curves show the theoretical visibilities expected
from uniform disks with diameters of 52, 57 and 62\,mas. }
\end{figure}

\fi

Note that we expect the calibrator star $\gamma$~Ret (HR~1264; $V=4.5$;
M4~III) to be slightly resolved.  This star has no measured angular
diameter, but \citeone{O+H82} estimate a diameter of 11\,mas based on its
spectral type.  Using the $V-K$ calibration of \citeone{DiB93} gives
7.5\,mas.  Correcting the visibility curve of R~Dor for the non-zero size
of the calibrator star leads to a slight upwards revision in the diameter
by 0.5--1\,mas, depending on the actual diameter of \gRet.

\if\preprint1
	\begin{figure}
\centerline{
\psfig{figure=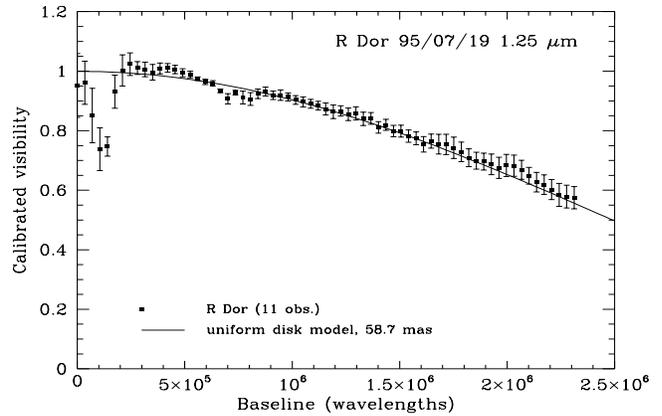%
,bbllx=61pt,bblly=55pt,bburx=443pt,bbury=302pt%
,width=\the\hsize,scale=100}}

\caption[]{\label{fig.vis-NTT1995} Visibilities of R~Dor at
1.25\,\micron\ with an annular mask on the NTT in 1995 July, showing the
average of 11 observations.  The solid curve shows the visibility expected
from a uniform disk with diameter 58.7\,mas. }
\end{figure}

\fi

We also observed R~Dor on 1993 August~6 using the $J$ filter, this time
with an annular mask that had an effective outer diameter of 3.3\,m and a
width of 20\,cm (lower panel of \figref{fig.vis-NTT1993}).  Further
observations using the same setup were obtained almost two years later, on
1995 July~19 (\figref{fig.vis-NTT1995}).  The former data set gave a
uniform-disk diameter of $(57\pm5)$\,mas and the latter gave $(59 \pm
3)$\,mas, both in good agreement with the results from the 7-hole mask.
The anomalous feature seen at short baselines in \figref{fig.vis-NTT1995}
is due to miscalibration caused by slight differences in seeing between
observations of the object and its calibrator star \cite{H+B92}.  The
effect is seen to lesser degrees in all of the plots presented here, and
the affected data were excluded from the fits to the visibility function.

\if\preprint1
	\begin{figure}
\centerline{
\psfig{figure=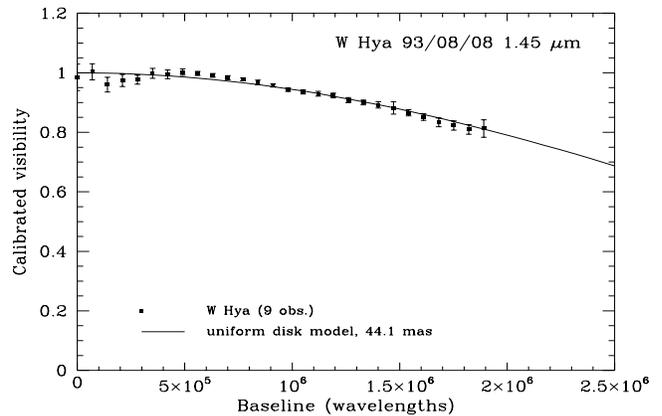%
,bbllx=61pt,bblly=55pt,bburx=443pt,bbury=302pt%
,width=\the\hsize,scale=100}}

\caption[]{\label{fig.vis-WHya} Visibilities of W~Hya measured at
1.45\,\micron\ with an annular mask on the NTT in 1993 August, based on the
average of 9 observations.  The scatter at short baselines is due to seeing
variations between observations of the target and the calibrator stars.
The solid curve shows the visibility expected from a uniform disk with
diameter 44.1\,mas.  }

\end{figure}

\fi

We observed W~Hya with the annular mask on 1993 August~8.  These
observations used a variable filter (CVF) with a resolving power
($\lambda/\Delta\lambda$) of about 50, set to a wavelength of
1.45\,\micron.  Two calibrator stars were used: HR~5192 and HR~5287.  The
result, shown in \figref{fig.vis-WHya}, is a uniform-disk diameter of
$44\pm4$\,mas.

\subsection{AAT Observations}

On 1995 January~13 we observed R~Dor and \aOri\ at 855\,nm with a slit mask
using the 3.9-m Anglo-Australian Telescope (AAT).  We used the MAPPIT
facility (Masked APerture-Plane Interference Telescope; \citebare{BRM94b}),
which consists of optical elements mounted on fixed rails at the \coude\
focus.  The system differed in two important respects from that described
by Bedding et al.  Firstly, the wavelength-dispersed system was not used.
Instead, the prism and cylindrical lens were replaced by an interference
filter ($\lambda=855$\,nm, $\Delta\lambda=40$\,nm).  Secondly, the detector
was a CCD with on-chip binning, similar to the system used by
\citeone{BHB90}.

For each observation we obtained 10\,000 short (13\,ms) exposures of the
target followed by an identical number of the calibrator star.  The
calibrator for R~Dor was again \gRet, while for \aOri\ we used \gOri.  The
aperture mask was a narrow slit whose effective width was 8\,cm.  The slit
was aligned diametrically so that the maximum baseline corresponded to the
full 3.89\,m aperture of the AAT\@.  The central 1.51\,m was obscured by
the secondary mirror, which results in a gap in spatial-frequency coverage:
baselines having lengths between 1.19\,m and 1.51\,m are not sampled.

\if\preprint1
	\begin{figure}
\centerline{
\psfig{figure=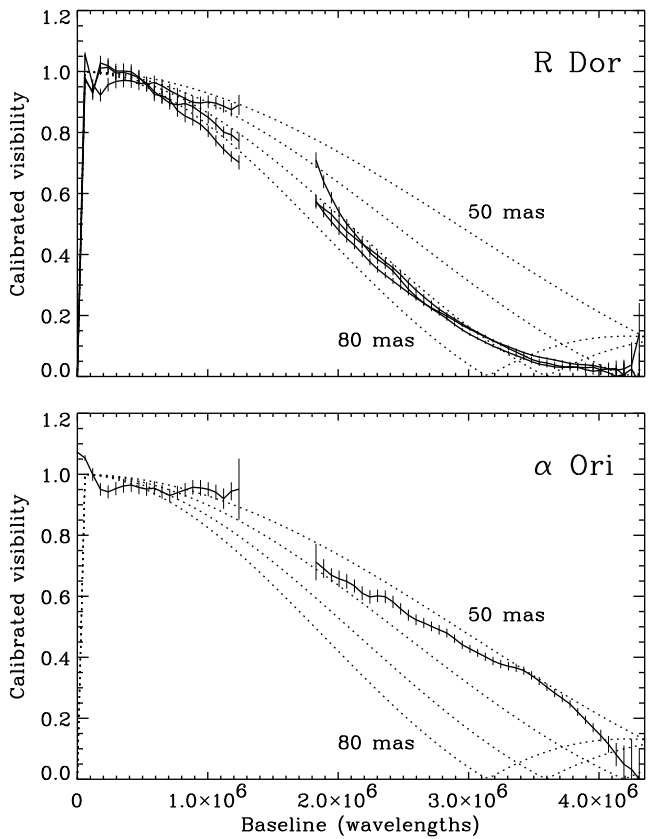%
,bbllx=114pt,bblly=31pt,bburx=516pt,bbury=555pt%
,width=\the\hsize,scale=100}}

\caption[]{\label{fig.vis-AAT} Visibilities of R~Dor (three observations,
upper panel) and \aOri\ (one observation, lower panel) measured at 855\,nm
with a slit mask on the AAT in 1995 January.  For comparison, the dotted
curves show the theoretical visiblities expected from uniform disks with
diameters ranging from 50\,mas (top) to 80\,mas (bottom) in steps of
10\,mas. }
\end{figure}

\fi

For R~Dor we obtained two observations with the slit oriented at position
angle 65\deg\ on the sky and a further observation at 145\deg.
\Figref{fig.vis-AAT} shows calibrated visiblities for these three
observations, as well as for a single observation of \aOri\ obtained at
position angle 115\deg.  The visibilities of R~Dor have been corrected for
the non-zero angular size of the calibrator star, assumed to be 10\,mas.
This correction is only a few percent, even at the longest baselines.  Note
that fringe visibilities from \aOri\ are likely to be affected by the
presence of surface features \cite{BHB90,WBB92}.  However, our aim here is
to establish that R~Dor exceeds \aOri\ in angular size, which is
demonstrated by our observations, at least at this wavelength.  We should
note that our measured angular diameters will be affected by limb darkening
and the presence of hotspots and non-circularity, which may be different
for the two stars.  By calculating the bispectrum of each observation, we
find that R~Dor shows non-zero closure phases
(\figref{fig.closure-phases}).  These imply an asymmetrical brightness
distribution similar to that previously found for the supergiant \aOri\ and
Mira variables such as \oCet\ \cite{WBB92,HGG92,THB94a}.  R~Dor is the
first non-Mira giant from which non-zero closure phases have been detected.
Note that \citeone{DiB+B90} found anomalously high visibilities on long
baselines for the star $\beta$~And (M0~III) which may indicate surface
structure, but those measurements did not provide phase information.

The limited position-angle coverage of our data prevents image
reconstruction and R~Dor is clearly a good candidate for more detailed
observations, particularly in view of its large angular size.

\section{Discussion}

\subsection{Angular diameter}

To obtain the true photospheric angular diameter of an M-star from a
measured diameter requires two corrections, both of which are strongly
dependent on wavelength.  The first correction is for limb darkening, which
makes a star appear smaller.  The second correction is for opacity effects
from deep molecular absorption bands, in which the star appears larger.
This latter effect explains why the angular size at 855\,nm is larger than
at $J$ band: the 855\,nm filter (with 40\,nm bandpass) contains a strong
TiO absorption feature whose depth is about 50\% of the continuum (see
Fig.~9 of \citebare{BBS89a}).

In the near infrared both effects are greatly reduced and the measured
diameter is within 1--2\% of the `true' stellar diameter
(\citebare{BBS89a}, \citeyear{BSW96}), where the latter is defined by the
point at which the Rosseland optical depth is unity.  We therefore adopt
our measurement of R~Dor at 1.25\,\micron\ ($57\pm5$\,mas) as the true
angular diameter, to be used in the next section.

Measurements at visible wavelengths can be corrected for limb-darkening and
opacity effects, although this correction is somewhat {\it ad hoc\/} due to
the lack of reliable model atmospheres.  \citeone{HST95} made these
corrections to their observations of W~Hya, which were obtained with a slit
mask on the WHT at wavelengths of 700 and 710\,nm.  They arrived at a
photospheric diameter of $46\pm6$\,mas (after corrections of 30--40\%).
This result agrees with our 1.45\,\micron\ measurement of $44\pm4$\,mas,
giving support to the accuracy of their correction process.

For comparison, the uniform-disk diameter of \aOri\ at 2.2\micron\ has been
measured to be 44\,mas \cite{DBR92}.  Thus, while R~Dor is the largest star
in the sky, W~Hya rivals \aOri\ for second place.

\subsection{Bolometric magnitude and effective temperature}

The effective temperature of a star can be determined from its angular
diameter $\phi$ and its apparent flux \cite{RJW80,BBS89a}:
\begin{equation}
 \log(\Teff/{\rm K}) = 4.22 - 0.1 m_{\rm bol} - 0.5 \log(\phi/{\rm mas}),
\end{equation}
where $m_{\rm bol}$ is the bolometric magnitude.

Several infrared magnitude determinations for R~Dor are available in the
literature \cite{NSW71,K+H94}, which we have converted to the Carter system
\cite{Car90,McG94}.  These are supplemented by several new observations
made using the 0.75-m telescope and the MKII Infrared Photometer at
Sutherland Observatory, South Africa.  To allow observation of such a
bright star, an aperture mask was placed over the telescope for some of the
observations.  The accuracy of these measurements is about 0.03 magnitudes
at $JHK$ and 0.05 at $L$ on the Carter system

We derive the following average magnitudes: $J=-2.51$, $H=-3.50$, $K=-3.91$
and $L=-4.30$.  The scatter between measurements at different epochs is
about 0.1 to 0.2 magnitudes, presumably reflecting the variability of the
star.  In addition, we have the 12 and 25 micron fluxes from the IRAS Point
Source Catalogue, which are 5160 and 1590\,Jy.  From these six wavebands we
estimate the average bolometric magnitude of R~Dor to be $-0.89 \pm 0.1$
using a spline-fitting program kindly provided by Dr. Whitelock.  This
ignores the stellar absorption bands in the near-infrared and may therefore
overestimate the brightness by up to 0.1 magnitude.

We also observed R~Dor in two bands at Sutherland on 1993 August~7,
contemporaneously with our infrared diameter measurement.  We obtained
$J=-2.59$ and $L=-4.37$, implying that R~Dor was about 0.07 magnitudes
brighter than the average values given above.  On this basis, we take the
bolometric magnitude at the epoch of our diameter measurement to be
$-0.96$, which yields an effective temperature of $2740 \pm 190$\,K\@.
This is somewhat higher than previous indirect estimates: \citeone{NSW71}
gave 2400\,K, \citeone{deJNH88} gave 2365\,K (after correcting for an
obvious misprint) and \citeone{J+S91} gave 2230\,K.  The differences may be
due to the inadequacies of the indirect methods used to estimate
temperatures from colour of such red stars, and also to the intrinsic
variability of R~Dor.

\subsection{Classification, distance and radius}

R~Doradus is an M8 giant and is the brightest, and presumably closest, star
with such a late spectral type \cite{Win71}.  The late spectral type makes
it likely that R~Dor lies on the asymptotic giant branch (AGB) rather than
the red giant branch.  The star is catalogued by \citeone{Kho88} as a
semiregular variable of type SRb with a period of 338 days.  Despite this
classification as semiregular, R~Dor is in many ways closer to the Miras
than to other SRb stars.  Its period is near the peak of the Mira period
distribution function (250--350 days), while SRb stars almost always have
much shorter periods (e.g., \citebare{K+H92}; \citebare{J+K92b}).  Its late
spectral type and large $J-K$ are also more typical of Miras than other
M-type stars \citeeg{Fea96}.

Further evidence for the Mira-like nature of R~Dor comes from evidence for
mass loss, which is commonly seen in Miras with periods of more than 300
days \cite{J+K92a}.  In R~Dor, evidence for mass loss comes from the
12-micron IRAS flux, which exceeds the ground-based $N$-band flux by a
factor of two, indicating that there may be extended emission.  There may
also be an extended component at 60 and 100 micron, as claimed by
\citeone{YPK93}.  In addition, the IRAS LRS spectrum \cite{V+C89} appears
to show a weak silicate feature, indicating recent mass loss.

Assuming that R~Dor is closely related to the Mira variables, as seems
likely from the preceding discussion, we can apply the period--luminosity
relations for Miras in the LMC given by \citeone{Fea96}.  For this we use
the average magnitudes obtained above.  We obtain a luminosity for R~Dor of
$6500\pm1400$\,\Lsol\ and a distance of $61\pm7$\,pc, with the result being
the same whether we use the period--bolometric magnitude or the
period--$K$-magnitude relation.  The uncertainty is based on the scatter in
the observed period--luminosity relation and the uncertainty in the
distance to the LMC\@.  Our distance for R~Dor agrees with estimates of
60\,pc by \citeone{J+S91} and 51\,pc by \citeone{Cel95}.

Comparison with the Hipparcos distance will soon be possible.  Being one of
the few red giants accessible to Hipparcos, probably closer even than
\oCet, R~Dor will be valuable in establishing the zero-point of the Mira
period--luminosity relation.  We note, however, that the accuracy of
Hipparcos parallaxes for these stars may be compromised by time-varying
surface features which, for the closest objects, may affect the centroid at
a level of a few mas.

This distance together with our observed angular diameter implies a stellar
radius of $370\pm50$\,\Rsol.  From the pulsation equation $Q =
P(M/\Msol)^{1/2} (R/\Rsol)^{-3/2}$ and assuming $Q=0.04$\,days (appropriate
for first overtone pulsation), we derive a mass of $0.7\pm0.3$\,\Msol.  All
the derived parameters are consistent with a classification of this star as
Mira-like, with the effective temperature being slightly higher than the
average for Miras \cite{HST95,Fea96}.

Although angular diameter measurements of Miras have favoured first
overtone pulsation \cite{THB94b,HST95}, there is recent evidence from
cluster long-period variables in the LMC that the dominant mode is the
fundamental \cite{W+S96}.  If we assume that R~Dor pulsates in the
fundamental mode, which means adopting a value of $Q=0.09$\,days
\cite{F+W82}, we obtain a mass of $3.5\pm1$\,\Msol.  This is higher than
expected from its period \cite{Fea89,V+W93}, giving further support to
first-overtone pulsation.

All previous measurements of the radii of Miras fall in the range
400--500\,\Rsol, which is taken by \citeone{HST95} as evidence that Miras
are associated with a well-defined instability strip.  The fact that R~Dor
shows a more irregular pulsation behaviour but with many characteristics of
a Mira is consistent with it lying near the edge of such a strip.

\section*{Acknowledgments}

We thank the staff at the AAT and NTT for their assistance.  We also thank
Reiner Hofmann for making the SHARP fore-optics, Gerardo Ihle and the staff
in the La Silla workshop for making the NTT masks, SHARP experts Lowell
Tacconi-Garman and Andreas Eckart for excellent support during the
observations and Andreas Quirrenbach for making NTT time available in 1995
for observations of R~Dor.  We are grateful to Dave Laney, Freddy Marang
and Patricia Whitelock for obtaining photometry at SAAO and we thank Mike
Bessell and Lawrence Cram for useful discussions.  We also thank the
referee, Chris Haniff, for many valuable suggestions.  The development of
MAPPIT was supported by a grant under the CSIRO Collaborative Program in
Information Technology, and by funds from the University of Sydney Research
Grants Scheme and the Australian Research Council.

\if\preprint0
	\clearpage
        \begin{figure}
\centerline{
\psfig{figure=figure-rdor-ntt.ps%
,bbllx=93pt,bblly=207pt,bburx=517pt,bbury=726pt%
,width=\the\hsize,scale=100}}

\caption[]{\label{fig.vis-NTT1993} Visibilities of R~Dor measured at
1.25\,\micron\ with the NTT in 1993 August.  The upper panel is for the
7-hole mask, with crosses showing visibilities on 21 baselines.  The lower
panel is for the annular mask, with data points being azimuthal averages.
In both panels the solid curves show the theoretical visibilities expected
from uniform disks with diameters of 52, 57 and 62\,mas. }
\end{figure}

        \begin{figure}
\centerline{
\psfig{figure=oskar-rdor.ps%
,bbllx=61pt,bblly=55pt,bburx=443pt,bbury=302pt%
,width=\the\hsize,scale=100}}

\caption[]{\label{fig.vis-NTT1995} Visibilities of R~Dor at
1.25\,\micron\ with an annular mask on the NTT in 1995 July, showing the
average of 11 observations.  The solid curve shows the visibility expected
from a uniform disk with diameter 58.7\,mas. }
\end{figure}

        \begin{figure}
\centerline{
\psfig{figure=oskar-whya.ps%
,bbllx=61pt,bblly=55pt,bburx=443pt,bbury=302pt%
,width=\the\hsize,scale=100}}

\caption[]{\label{fig.vis-WHya} Visibilities of W~Hya measured at
1.45\,\micron\ with an annular mask on the NTT in 1993 August, based on the
average of 9 observations.  The scatter at short baselines is due to seeing
variations between observations of the target and the calibrator stars.
The solid curve shows the visibility expected from a uniform disk with
diameter 44.1\,mas.  }

\end{figure}

        \begin{figure}
\centerline{
\psfig{figure=rdor-mappit.ps%
,bbllx=114pt,bblly=31pt,bburx=516pt,bbury=555pt%
,width=\the\hsize,scale=100}}

\caption[]{\label{fig.vis-AAT} Visibilities of R~Dor (three observations,
upper panel) and \aOri\ (one observation, lower panel) measured at 855\,nm
with a slit mask on the AAT in 1995 January.  For comparison, the dotted
curves show the theoretical visiblities expected from uniform disks with
diameters ranging from 50\,mas (top) to 80\,mas (bottom) in steps of
10\,mas. }
\end{figure}

	\if\preprint1
 \onecolumn
\fi
\begin{figure}
\centerline{
\psfig{figure=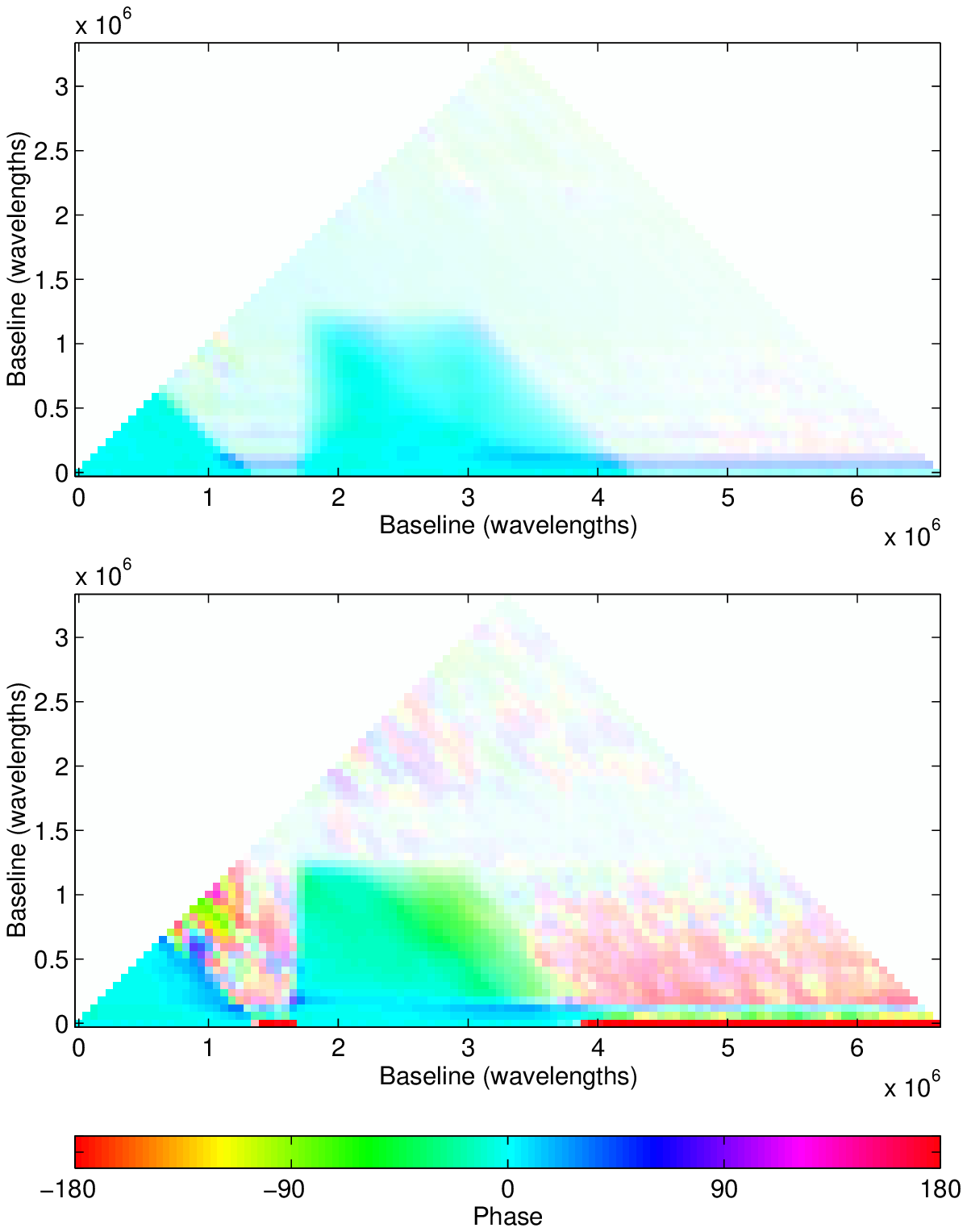%
,bbllx=136pt,bblly=268pt,bburx=481pt,bbury=706pt%
,width=\the\hsize,scale=80}}

\if\preprint1
{\sf  This colour figure is available from
{\tt ftp.physics.usyd.edu.au} in {\tt /pub/tmp/bedding/closure.ps}}\\
\fi
\caption[]{\label{fig.closure-phases} Bispectra from observations of R~Dor
(lower) and the calibrator \gRet\ (upper) using a slit mask on the AAT\@.
The amplitude and phase of the complex bispectrum at each point are
represented by intensity and colour, respectively, with blue corresponding
to zero closure phase.  The axes are spatial frequencies along the slit
($u_1$ and $u_2$), so that each point in the plane defines a triangle of
spatial frequencies ($u_1$, $u_2$ and $u_1+u_2$).  Only the nondegenerate
portion of the bispectrum is shown here, which is contained in a triangle
bounded by the lines $u_1=0$, $u_1=u_2$ and $u_1+u_2 = u_{\rm max}$, where
$u_{\rm max}$ is the maximum spatial frequency sampled by the slit (see
\citebare{B+H93}).  The green region in the bispectrum of R~Dor at long
baselines (high spatial frequencies) indicates non-zero closure phases.

}
\end{figure}
\psfull

\else
	\if\preprint1
 \onecolumn
\fi
\begin{figure}
\centerline{
\psfig{figure=figure-closure.ps%
,bbllx=136pt,bblly=268pt,bburx=481pt,bbury=706pt%
,width=\the\hsize,scale=80}}

\if\preprint1
{\sf  This colour figure is available from
{\tt ftp.physics.usyd.edu.au} in {\tt /pub/tmp/bedding/closure.ps}}\\
\fi
\caption[]{\label{fig.closure-phases} Bispectra from observations of R~Dor
(lower) and the calibrator \gRet\ (upper) using a slit mask on the AAT\@.
The amplitude and phase of the complex bispectrum at each point are
represented by intensity and colour, respectively, with blue corresponding
to zero closure phase.  The axes are spatial frequencies along the slit
($u_1$ and $u_2$), so that each point in the plane defines a triangle of
spatial frequencies ($u_1$, $u_2$ and $u_1+u_2$).  Only the nondegenerate
portion of the bispectrum is shown here, which is contained in a triangle
bounded by the lines $u_1=0$, $u_1=u_2$ and $u_1+u_2 = u_{\rm max}$, where
$u_{\rm max}$ is the maximum spatial frequency sampled by the slit (see
\citebare{B+H93}).  The green region in the bispectrum of R~Dor at long
baselines (high spatial frequencies) indicates non-zero closure phases.

}
\end{figure}
\psfull

	\bsp
\fi

\end{document}